\documentclass[]{optica-article}
\journal{opticajournal} 

\usepackage{array}
\usepackage{afterpage}
\usepackage{siunitx}

\usepackage{ulem}
\usepackage{multirow}
\usepackage{xcolor} 
\definecolor{colour1}{RGB}{232,232,232} 
\definecolor{colour2}{RGB}{215,244,245} 

\usepackage[bottom]{footmisc}

\usepackage{braket}

\usepackage[acronym]{glossaries}

\newacronym{qi}{QI}{quantum information}
\newacronym{qc}{QC}{quantum communications}
\newacronym{qkd}{QKD}{quantum key distribution}

\newacronym{df}{DF}{dark fibre}
\newacronym{cd}{CD}{chromatic dispersion}
\newacronym{pmd}{PMD}{polarisation mode dispersion}
\newacronym{fwhm}{FWHM}{full width at half maximum}
\newacronym{smf}{SMF}{single mode fibre}
\newacronym{nzdsf}{NZDSF}{non-zero dispersion shifted fibre}
\newacronym{car}{CAR}{coincidence-to-accidental ratio}

\newacronym{Tq}{Tq}{quantum transmitter}
\newacronym{Rq}{Rq}{quantum receiver}
\newacronym{skr}{SKR}{secret key rate}
\newacronym{ppr}{PPR}{photon-pair production rate}
\newacronym{qber}{QBER}{quantum bit error rate}

\newacronym{tbp}{TBP}{Time-Bandwidth product}
\newacronym{wdm}{WDM}{wavelength division multiplexing} 

\usepackage{pdflscape}

\usepackage{lineno}

\begin{document}

\title{
Optimising physical parameters of a quantum network based on a loss-jitter trade-off
}

\author{
Marcus~J.~Clark\authormark{1,$\dagger$ }    
and Siddarth~K.~Joshi\authormark{1,*} 
}

\address{\authormark{1} Quantum Engineering Technology Labs \& School of Electrical, Electronic, and Mechanical Engineering, University of Bristol, Tyndall Avenue, Bristol, BS8 1FD UK\\
}

\email{
\authormark{$\dagger$}mj.clark@bristol.ac.uk, 
\authormark{*}sk.joshi@bristol.ac.uk
} 

\begin{abstract*}
As quantum communication systems and networks are becoming a commercial reality, clarity on their future infrastructure is increasingly important. 
Based on the inevitable presence of some amount of loss, chromatic dispersion, and timing jitter, we present simulations to show that certain wavelengths and bandwidths have clear advantages. 
\end{abstract*}

\section{Introduction}\label{Sec:intro}

The purpose of a quantum network is to distribute \gls{qi}. 
Such networks are currently deployed for the application of \gls{qkd}~\cite{cao2022evolution, liu2022towards}, but advancements in entanglement distribution~\cite{Caleffi2024DistributedGates,Serafini2006distributed,main2025distributed} demonstrate their use in \gls{qi} processing tasks. 
The underlying principles of no-cloning and monogamy of entanglement meant that quantum signals cannot be amplified like classical ones. 
Similarly, the use of a relay compromises security and prevents end-to-end distribution of \gls{qi}. 
Further, since all \gls{qi} tasks are highly susceptible to noise, temporal filtering during pre/post-selection of events is the most common way of improving the signal to noise ratio~\cite{Bouchard2021achieving,Christandl2009postselection,Pan2024evolution}. 
Practically, the only limitations to this temporal filtering (or gating) is the timing jitter of the electronics/detectors used, combined with the temporal dispersion and coherence time of the transmitted qubits (directly related to their optical bandwidth).
Thus, quantum technologies benefit from both low-loss and low-dispersion networks.
Practically, there is a trade-off between the loss and dispersion, where one is prioritised.

In typical deployed fibre based \gls{qkd} systems, the wavelength normally falls into either the low-loss optical band (C-Band) or the low-dispersion (i.e., high temporal distinguishability) optical band (O-Band).
The decision of which system to deploy across different types of infrastructure is further complicated by economic factors, the availability of the infrastructure, and losses in network infrastructure (such as filters, optical switches, multiplexing devices, available quantum emitters, etc.).
For example, the choice of Craddock et.~al.~\cite{Qunnect2024} to use the O-Band is driven by the extreme costs of new fibre and the need to piggyback quantum signals onto existing telecommunications infrastructure while using wavelengths that are native to quantum systems. 
While the same optical band was chosen by Beumann et.~al.~\cite{Neumann2021} to exploit the wavelength entanglement of photons above and below the zero-dispersion wavelength to maximise temporal distinguishability.
However, most quantum communications demonstrations~\cite{sibson2017chip, zheng2023multichip, jing2024experimental} use the C-Band to avoid the larger attenuation at the O-Band.
Furthermore, the C-Band allows access to standards defining multiple, compact, wavelength channels allowing research groups to exploit these for networking~\cite{wengerowsky2018entanglement,joshi2020trusted, qi2021ANetwork, wen2022realizing}.

Many recent implementations of quantum networking have changed to O-Band photons for economic reasons.
This allows coexistence of quantum and classical signals in a single optical fibre with wavelengths separated by a large spectral gap ($1360$\,nm to $1530$\,nm)~\cite{thomas2023designing,thomas2023optimization,mehdizadeh2024quantum,Chung2022Illinois}.
Such schemes reduce the cost of implementation when compared to a situation where quantum and classical signals require independent optical fibres.
Concurrently, classical communications research has begun to explore extreme wideband communications multiplexing, leading to simultaneous transmission of classical communications signals across the whole telecommunications spectrum, specifically the OESCLU-Bands ($1260$\,nm to $1626$\,nm)~\cite{puttnam2025HighTransmission}.
This would prevent coexistence of quantum and classical communications in the same fibre due to scattered noise effects from neighbouring wavelengths.
As such,  this work will focus on quantum communications with no coexistence of quantum and classical channels. 
Further discussion on coexistence can be found in Supplement 1 Section 1.

As optical fibre infrastructure continually improves, constraints applied by the type of fibre used become less impactful.
However, the constraints of loss, bandwidth, and temporal considerations are fundamental for all optical transmission regardless of any current or future medium used. 
Thus, we present baseline simulation looking at the trade-offs between bandwidth, dispersion, loss and coherence length, valid for a wide class of qubits.
Using our experimentally calibrated simulations, we present a simple process to determine the optimal wavelength and bandwidth to use in a quantum network knowing the values for loss, dispersion, and detection jitter present in any given system. 
Our results show how to choose the optimum wavelength and bandwidth for any set of fibre parameters (i.e., length, loss and dispersion), for a given optical fibre. 
This is most useful when the quantum network must use existing fibre infrastructure. 
Here, we consider both standard telecommunications single mode fibre (SMF)~\cite{CorningUltra} and one example of non-zero dispersion shifted fibre (NZDSF)~\cite{CorningLEAF}, finding the optimal combination of wavelength, bandwidth, and fibre for each distance-jitter case. 
Lastly, we discuss the difference between optimising a single channel and a wavelength multiplexed network.

We consider a discrete variable entanglement distribution scheme implementing the BBM92 protocol~\cite{BBM92} for Quantum Key Distribution. 
Nevertheless, our methodology is sufficiently general to be adapted to any distributed \gls{qi} processing task.

\section{Methodology}\label{Sec:method}

\begin{figure}[!t]
    \centering
    \includegraphics[width=10cm]{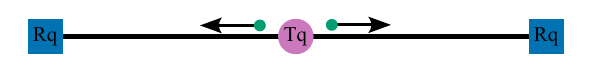}
    \caption{
    The schematic for \gls{qi} sharing in a source central entanglement distribution scheme. 
    Here \gls{Tq} is any source of \gls{qi} or entanglement, and \gls{Rq} is any system that can receive \gls{qi} or perform quantum state measurements.
    }
    \label{fig:QIonly}
\end{figure}

To simulate a quantum information sharing link, one must first define the architecture of the system sharing the information.
This work uses a central source of entanglement for the network, as shown in Fig.~\ref{fig:QIonly}.
The source is defined to produce pairs of entangled photons, where we take the entanglement to be in both the time-energy domain and the polarisation domain.
Networks with such entanglement sharing schemes were demonstrated with many simultaneous users by Wengerowsky et. al.~\cite{wengerowsky2018entanglement}.
Here we simulate single links from an entangled photon pair source at a central source, to two remote user nodes.
As such there are many parameters that effect the transmitted entangled photons before measurement.

As well as considering effects from attenuation, \gls{cd}, jitter, and time-bandwidth uncertainty, many other effects are considered.
The noise mechanisms included are detector dark counts, single photons from pairs where one photon is lost, and additional photons from wavelength or temporal multiplexing.
Additionally to the link loss, the additional loss from temporal multiplexing, dispersion compensation devices, quantum receiver transmission efficiency, and single photon detector efficiency.
The temporal width of the measurement is an important factor in the simulator, which is comprised of contributions from detector jitter, time synchronisation jitter, \gls{cd}, and time-bandwidth uncertainty.
The tool does not consider \gls{pmd}, as the contribution is negligible in most situation, other than when there is a negligible \gls{cd} and large bandwidth. 
In this case it would impact the results by overestimation of link performance.

Detailed discussion on the simulation tool can be found in the Supplement 1 Section 2.
This tool was tested against experimental data from Clark et. al.~\cite{clark2023entanglement} and other such prior work\footnote{Full list of prior work which the simulator was tested against can be found in the Supplement 1 Section 2.} showing the accuracy of the tool for simulating the links.

\section{Results}\label{Sec:results}

An optical signal travelling through a medium experiences \gls{cd} and attenuation. 
In the case of entangled photons, detected using single-photon detectors, we must measure the arrival time of the photon and correlate that with the arrival time of its twin. 
There are two more sources of temporal pulse broadening: the detection jitter and the time synchronisation jitter, together referred to as the total measurement jitter. 
For the quantum information distributed by pairs of entangled photons to be useful, we must distinguish correlated pairs of photons (the information carriers) from the uncorrelated single photons (noise) that do not carry meaningful quantum information. 
One major source of this noise is photon pairs where one photon has been lost. 
Similarly, the main method of filtering out noise is using temporal correlations. 
These noise production, suppression, and filtering mechanisms naturally have a trade-off. 
These are not the only processes contributing to the Signal or the Noise, but these are the most basic and inevitable in any experimental system. 
Thus, it is meaningful to study the trade-off between the signal attenuation and the temporal widening of the entanglement distribution.

Lastly, it is difficult to quantify the amount of "useful quantum information" distributed without specifying a protocol. 
We have chosen the BBM92 protocol~\cite{BBM92} for \gls{qkd} because it is very well studied and our simulator is calibrated against experimental results for this protocol. 
Thus, the \gls{skr} is an excellent stand-in for the amount of useful quantum information distributed.

To study the effect of the fibre length, measurement jitter, and the bandwidth we use optimisation mosaics to distil many heatmaps of SKR into a single discrete mosaic showing the most optimal bandwidth for each transmission distance and jitter.
We can the further condense this to find the optimal selection of optical fibre, wavelength, and bandwidth for each scenario.
Full details on how these mosaics are constructed can be found in the Supplement 1 Section 3.

\subsection{The optimal bandwidth}\label{subSec:bandwidth}

\begin{figure}[!t]
    \centering
    \includegraphics[width=12.5cm]{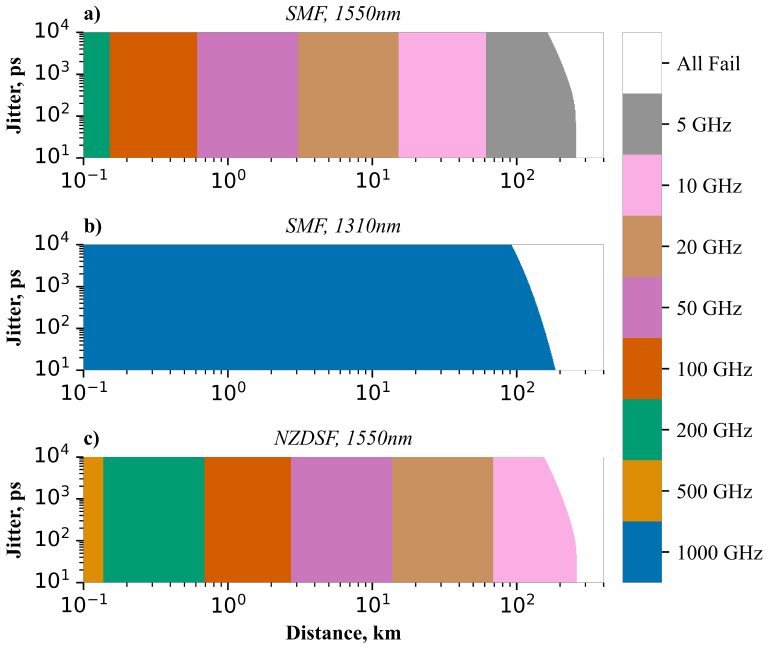} 
    \caption{
    The choice of bandwidth of transmitted photon that resulted in the highest secret key rate, against total measurement jitter and the transmission distance.
    \textbf{a)} shows the optimal selection for SMF fibre with \SI{1550}{\nano\meter} light.
    \textbf{b)} shows the optimal selection for SMF fibre with \SI{1310}{\nano\meter} light.
    \textbf{c)} shows the optimal selection for NZDSF fibre with \SI{1550}{\nano\meter} light.
    Here SMF is Corning SMF-28Ultra~\cite{CorningUltra} and NZDSF is Corning LEAF~\cite{CorningLEAF}.
    }
    \label{fig:1ch_comparison_seperate}
\end{figure}

\begin{figure}[!b]
    \centering
    \includegraphics[width=13cm]{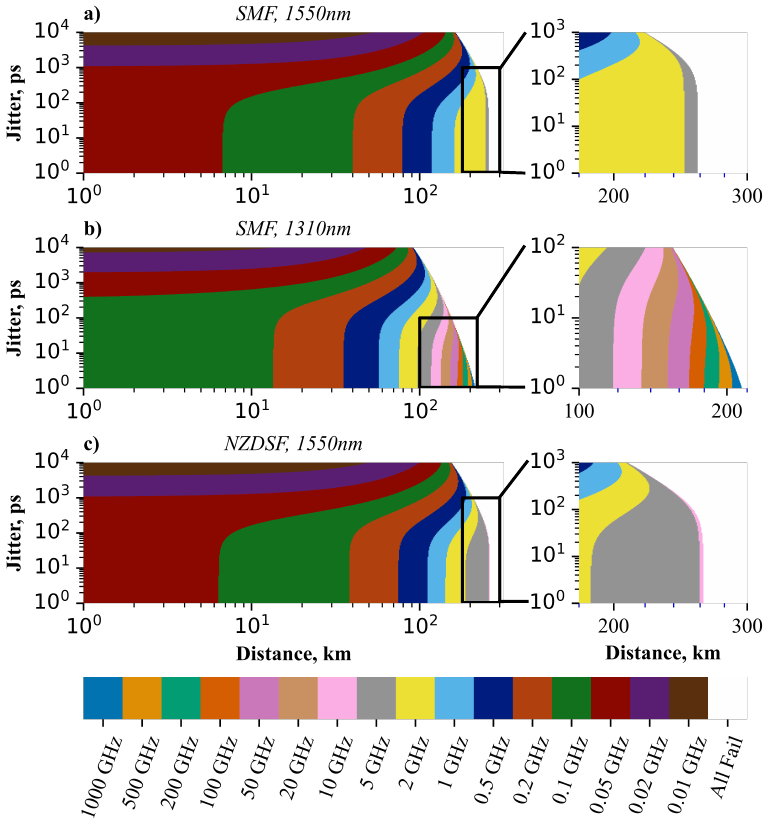}
    \caption{
    The choice of bandwidth of transmitted photon that resulted in the highest secret key rate per GHz of the bandwidth, against total measurement jitter and the transmission distance.
    \textbf{a)} shows SMF fibre with \SI{1550}{\nano\meter} light. 
    \textbf{b)} shows SMF fibre with \SI{1310}{\nano\meter} light.
    \textbf{c)} shows NZDSF fibre with \SI{1550}{\nano\meter} light.
    Here SMF is Corning SMF-28Ultra~\cite{CorningUltra} and NZDSF is Corning LEAF~\cite{CorningLEAF}.
    }
    \label{fig:Compair3Scenarios_Seperate_perGHz}
\end{figure}

While the ideal bandwidth in any practical sense can be governed by availability, costs, etc. an optimal value can be calculated from a theoretical perspective.
Broadband signals have a narrow temporal width due to its coherence time, but experiences higher \gls{cd}, leading to a trade-off between the two effects, as shown in Supplement 1 Fig.~S2.
In this case, as long as the total jitter is low, a bandwidth of about 10\,GHz is near optimal. 
This proves to be approximately true for wavelengths that experience a non-negligible \gls{cd} in standard telecoms grade optical fibre~\cite{CorningUltra,CorningLEAF}.
At larger bandwidths, high \gls{skr} can only be achieved for short fibre distances, due to the significant increase in \gls{cd}. 
If the bandwidth is narrow the effect of a long coherence time (e.g., $\approx$ \SI{0.4}{\nano\second} for \SI{1}{\giga\hertz} bandwidth) of each photon results in a significant effective jitter.
One can think of this as if the minimum effective total jitter bing limited by the coherence time.

The above is an oversimplification if we take into account the different types of optical fibre and separate the effects of \gls{cd} from the other effects contributing to jitter. 
Considering the time synchronisation and detector jitter together called measurement jitter, we can look at Fig.~\ref{fig:1ch_comparison_seperate} which compares the transmission of a \SI{1550}{\nano\meter} quantum signal through Single Mode Fibre Corning SMF-28 Ultra~\cite{CorningUltra}, \SI{1310}{\nano\meter} signal through the same fibre, and a \SI{1550}{\nano\meter} signal through a Non-Zero Dispersion Shifted Fibre Corning LEAF~\cite{CorningLEAF}.
From here, Single Mode Fibre Corning SMF-28 Ultra~\cite{CorningUltra} will be referred to as SMF, and Non-Zero Dispersion Shifted Fibre Corning LEAF~\cite{CorningLEAF} will be referred to as NZDSF.
We note that we have used these fibres and wavelengths for illustrative purposes, and the calculation can be repeated for any fibre wavelength combination.

Interestingly, in Fig~\ref{fig:1ch_comparison_seperate} \textbf{b)} we see that a bandwidth of \SI{1000}{\giga\hertz} is uniformly better across all distances for \SI{1310}{\nano\meter}. 
This is because the y-axis is the measurement jitter that does not include the effects of \gls{cd}. 
In our simulations 1310\,nm is taken to be the zero-dispersion wavelength of the fibre, defining the contribution of \gls{cd} as zero. 
So, the contribution from the coherence time dominate, favouring large bandwidths. 
This is also why shorter fibres with 1550\,nm favour slightly larger bandwidths. 
Similarly, for the NZDSF where the \gls{cd} is lower, the optimal bandwidth at each distance is larger. 
This is true despite the slightly higher loss of the NZDSF (\SI{0.19}{\decibel\per\kilo\meter}~\cite{CorningLEAF}) compared to the SMF (\SI{0.18}{\decibel\per\kilo\meter}~\cite{CorningUltra}).

All above calculations were performed assuming that the entire capacity of a dark fibre was dedicated to a single \gls{qkd} link. 
However, wavelength multiplexing of quantum networks has proven to be an excellent method to improve scalability of quantum networks, beyond single pairwise links~\cite{wengerowsky2018entanglement}. 
Thus we want to look at the total key rate obtainable (i.e. as an indicator of the total amount of useful quantum information that can be transferred) if as many multiplexed channels as possible were used. An easy way to look at this is to consider the total amount of secret key per GHz of linewidth used.

Fig~\ref{fig:Compair3Scenarios_Seperate_perGHz} compares the same three scenarios as above, but looks at the total \gls{skr} per GHz. 
Naturally, even smaller bandwidths prove to be optimal. 
This is because the marginal gain in key rate by using a larger bandwidth (i.e. smaller coherence time) is overshadowed by the ability to utilize more wavelength channels.

In the case of very high measurement jitter, the smallest bandwidths tend to universally perform better until the coherence time becomes comparable to the jitter. 
At which point, the next larger bandwidth step used in the simulation proves to be optimal. 
This repeats until the \gls{cd} imposed by high bandwidths becomes significant and prevents key generation. 
Likewise, for the case where we operate close to the zero-dispersion wavelength (e.g.,1310\,nm light in SMF 28 Ultra fibre), higher bandwidths (compared to Fig~\ref{fig:Compair3Scenarios_Seperate_perGHz}a and c) are useful at very long distances because of the reduced \gls{cd}.

We note that in all our calculations, we have assumed that the transmitter is using the maximum/optimum modulation/signal rate given the limitations of the receiver (i.e. the measurement jitter). 
This can easily be achieved for many degrees of freedom of entangled states including, polarisation, time-bin, spatial modes. etc. 
However, some modulation schemes have a limit to the modulation rate based on the available optical bandwidth and these results would need to be modified for those.
Additionally, we note that a real network is very likely to have varying distances between different sets of users. 
For any given pair of users where the source is not in the middle, the above analysis can be modified to minimise the trade-off on each link's performance.

\subsection{Optimum fibre, wavelength and bandwidth combination for a single link}

\begin{figure}[!b]
    \centering
    \includegraphics[width=13cm]{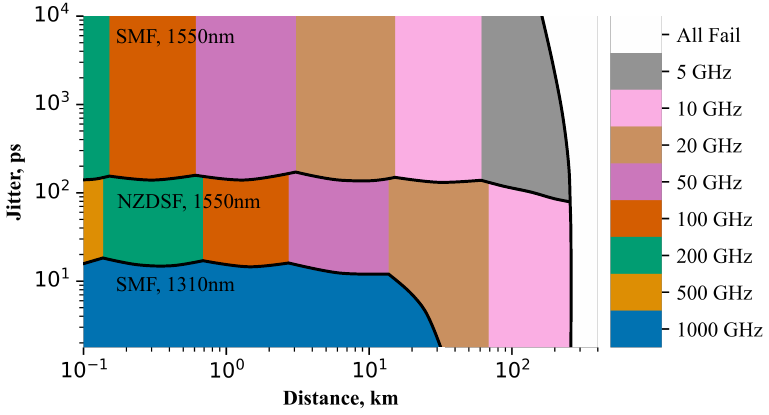}
    \caption{
    The combined selection of optimal transmission wavelength, transmission medium, and linewidth that resulted in the highest secret key rate, against total measurement jitter and the transmission distance.
    Here SMF is Corning SMF-28Ultra~\cite{CorningUltra} and NZDSF is Corning LEAF~\cite{CorningLEAF}.
    within the SMF \SI{1310}{\nano\meter} section, the two dotted lines show the region where there is an improvement over SMF and NZDSF at \SI{1550}{\nano\meter}, but where the performance is worse than SMF \SI{1310}{\nano\meter} at the higher bandwidth.
    }
    \label{fig:1ch_comparison_allin1}
\end{figure}

When only a single link is considered, we can find an optimal bandwidth, signal wavelength, and operation medium for any given combination of fibre distance and measurement jitter, as shown in Fig.~\ref{fig:1ch_comparison_allin1}.
Here, we see three distinct regions where different combinations of wavelength and optical fibre are optimal, and they are divided by the total measurement jitter.
When the measurement jitter is high (\SI{>100}{\pico\second}), the combination resulting in the lowest transmission loss is optimal, specifically a \SI{1550}{\nano\meter} signal in SMF.
With low jitters (\SI{<20}{\pico\second}) and limited attenuation (\SI{<30}{\kilo\meter}), an advantage can be found using a signal at the zero-dispersion wavelength, specifically with a \SI{1310}{\nano\meter} signal in SMF.
The intermediate region (between \SI{20}{\pico\second} and \SI{100}{\pico\second}), using NZDSF is seen to be optimal, with a trade-off between the \gls{cd} and attenuation.

Inside each region with non-zero \gls{cd}, there is a dependence on the propagation distance, as to which bandwidth is most optimal.
Specifically, as the distance increases the optimal bandwidth reduces.
This comes directly from the increase in \gls{cd} as the signal propagates, leading to more noise with larger bandwidths.
These bands are simply the bandwidths that equates to the lowest total signal broadening from the time-bandwidth product and \gls{cd}.

In this scenario, only one signal will be sent through the optical fibre.
As such, the optimal selection can be directly take from Fig.~\ref{fig:1ch_comparison_allin1}.
However, the performance of lower bandwidth signals that are optimal at long distances (\SI{\approx100}{\kilo\meter}) are similar in performance to the optimal selection of bandwidth at shorter distance (\SI{\approx10}{\kilo\meter}).
Specifically, when looking at \SI{10}{\kilo\meter} and \SI{200}{\pico\second} of jitter the optimal selection of bandwidth is \SI{20}{\giga\hertz}, but bandwidths in the range \qtyrange{5}{50}{\giga\hertz} recover at least \SI{90}{\%} of the optimal \gls{skr}, as can be seen in tab.~S1 and tab.~S2.
This then allows for the optimal selection at \SI{100}{\kilo\meter} and \SI{200}{\pico\second} of jitter (\SI{5}{\giga\hertz} bandwidth) to be used for shorter distances with only a \SI{\approx6}{\%} drop in the \gls{skr}.
Access to data can be found in the Data Availability section for full comparison between all scenarios.

\subsection{Optimum combination for wavelength multiplexed quantum networks}

\afterpage{
  \clearpage

\begin{figure}[!t]
    \centering
    \includegraphics[width=13cm]{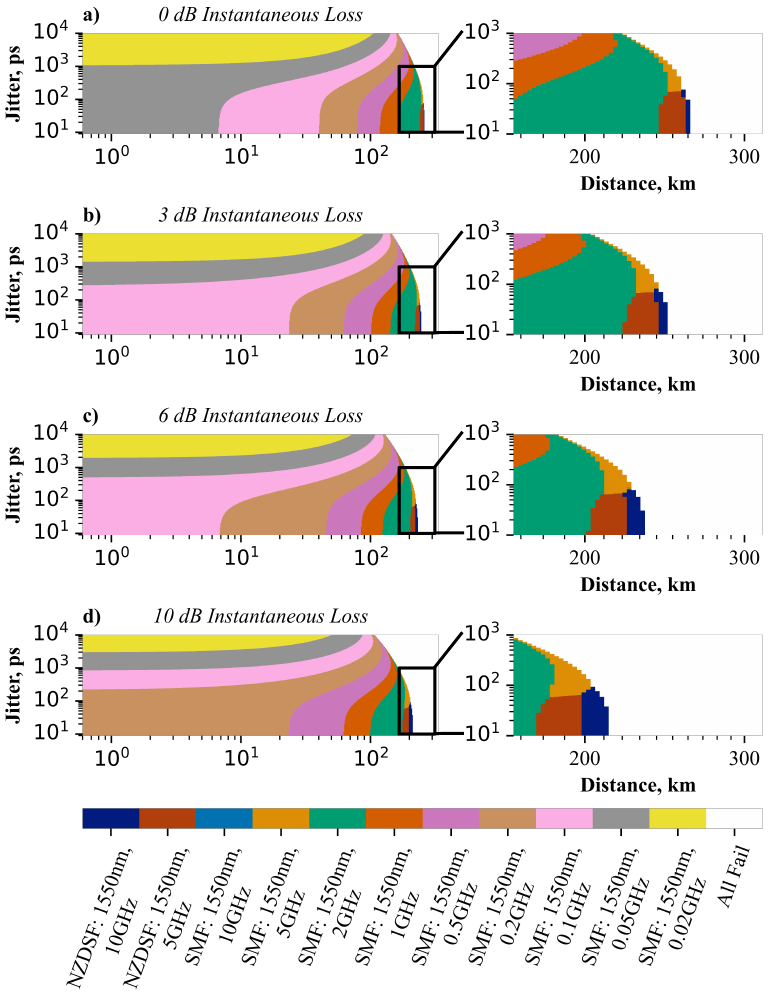}
    \caption{
    The combined selection of optimal transmission medium, transmission wavelength, and linewidth that resulted in the highest secret key rate per GHz of bandwidth, against total measurement jitter and the transmission distance.
    Figures \textbf{b)}-\textbf{d)} add instantaneous loss to the communication link.
    \textbf{a)} shows the situation with \SI{0}{dB} instantaneous loss. 
    \textbf{b)} shows the situation with \SI{3}{dB} instantaneous loss.
    \textbf{c)} shows the situation with \SI{6}{dB} instantaneous loss.
    \textbf{d)} shows the situation with \SI{10}{dB} instantaneous loss.
    Here SMF is Corning SMF-28Ultra~\cite{CorningUltra} and NZDSF is Corning LEAF~\cite{CorningLEAF}.
    Note that SMF at $1310$\,nm does not appear.
    }
    \label{fig:InstantLoss_Together_perGHz}
\end{figure}

\clearpage
}

By now looking again at the \gls{skr} per GHz bandwidth, we can optimise over all fibre and wavelength combinations.
The pattern shown by this optimisation mosaic is more complex, as shown in Fig.~\ref{fig:InstantLoss_Together_perGHz} \textbf{a)}. 
Unlike in the single communication channel case, there are not bands of different fibre types and wavelengths being optimal.
We now see that SMF with a signal at \SI{1550}{\nano\meter} is the optimal for almost all situation, except at the furthest propagation distances (\SI{>400}{\kilo\meter}) and the lowest jitters (\SI{<60}{\pico\second}) where NZDSF becomes the optimal selection.
The optimal bandwidths for long distances (\SI{>10}{\kilo\meter}) in the per GHz case (Fig.~\ref{fig:InstantLoss_Together_perGHz} \textbf{a)}) are smaller than those shown in the single channel case (Fig.~\ref{fig:1ch_comparison_allin1}), where the ranges are \qtyrange{0.02}{5}{\giga\hertz} and \qtyrange{5}{50}{\giga\hertz} respectively.
However, there is an overlapping section, specifically between \qtyrange{5}{10}{\giga\hertz}.
This would suggest that a system with a single fixed bandwidth would perform most optimally, for an arbitrary situation, with a bandwidth of \SI{5}{\giga\hertz} in SMF at \SI{1550}{\nano\meter} and with a bandwidth of \SI{10}{\giga\hertz} in NZDSF at \SI{1550}{\nano\meter}.

Looking directly at the data, we can see that for short distances, the change over bandwidth remains the same for each fibre wavelength combination, meaning that all wavelengths share an optimal bandwidth.
Specifically, at \SI{10}{\kilo\meter} we can see that all are optimal between \SI{0.1}{\pico\second} and \SI{0.05}{\pico\second}, as shown in tab.~S4.
When looking at a longer distance link, \SI{100}{\kilo\meter}, we can see that SMF and NZDSF for \SI{1550}{\nano\meter} light share an optimal bandwidth between \SI{0.05}{\giga\hertz} and \SI{0.02}{\giga\hertz}, but where \SI{1310}{\nano\meter} in SMF starts to fall lower and recover an extremely reduced \gls{skr}, as shown in tab.~S3.

When wavelength multiplexing many channels, there is unlikely to be a scenario where SMF at \SI{1310}{\nano\meter} would provide an increase in the \gls{skr} over the alternatives, where dark fibre is used for transmission of quantum signals.
This work also assumes that we have wavelength multiplexing technology that matches performance in both the O-Band and the C-Band, which is not the case.
Current dense wavelength division multiplexing technology provides multiplexing with narrow bandwidths only provide low loss multiplexing in the C-Band.

\section{Conclusion}\label{Sec:conclusion}

This work demonstrates that optimisation of not only wavelength, but the bandwidth of signals, can improve the \gls{skr} of \gls{qkd} links.
However, there is no clear optimum choice of operational parameters, optimum choice of fibre, or optimal combination of both for all combinations of jitter and transmission distance.
In the two discussed situations, single signal and wavelength multiplexed signals, there are different optimal bandwidths and optical bands across the jitter and optimum distance domain.
For a single link with typical system measurement jitter ($>20$\,ps), the optical C-Band is optimal with transmission through both SMF and NZDSF.
In this situation the bandwidths is based on the total transmission distance of the link, going from $\approx 100$\,GHz at $1$\,km to $\approx 5$\,GHz at $100$\,km.
For low measurement jitter the optical O-Band then becomes optimal for short distance links ($<20$\,km) transmission distances but only with large bandwidths ($>1000$\,GHz), and C-Band in NZDSF for longer distances ($>20$\,km) with bandwidths around $10$\,GHz.

For wavelength multiplexed links, the optical O-Band then becomes suboptimal across the bandwidth-distance domain, given assigning large transmission bandwidths where O-Band is optimal conflicts with fitting more channels into the optical link.
Across the bandwidth-distance domain, SMF is seen to be optimal, with the exception for the extreme transmission distance ($>180$\,km) case.
Here the optimal bandwidth is $5$\,GHz to $10$\,GHz, where this optimum is seen below $100$\,ps of total measurement jitter.
Across the domain, the highest optimal bandwidth is $5$\,GHz for the SMF case.
Small gains are seen by reducing the bandwidth as the tranmsission distance is reduced and the measurement jitter is increased.
For NZDSF the optimal selection is  $5$\,GHz to $10$\,GHz, 

For standard commercial use, where typical link lengths are $80$\,km%
\footnote{This distance is based on knowledge shared by UK based telecommunications researchers embedded with national operators.}, 
the maximum flexibility in terms of users situated at different lengths of fibre would be to transmit signals in the C-Band, where the ideal bandwidth would be $\approx 5$\,GHz for both NZDSF and SMF.
Improving the optimisation beyond this fixed multiplexing case, changing from fixed-bandwidth wavelength multiplexing technology to a dynamically tunable multiplexing technology would allow each user pair to have a tailored optical bandwidth, however only limited improvement would be seen.

We highlight the importance of optimising both wavelength and bandwidth of \gls{qc} devices.
This is especially true with commercial devices and presenting the operational parameters to customers such that they can perform their own channel optimisation.

See Supplement 1 for supporting content.

\begin{backmatter}

\bmsection{Author Contributions}
SJ produced the simulation tool used in this work.
MJC ran the simulations, collated all data, analysed the data, and wrote the manuscript and the supplemental document.
SJ provided supervision for the project and assistance in producing the manuscript and supplemental document.

\bmsection{Funding}
This work was supported by the UK Engineering and Physical Sciences Research Council (EPSRC) grants EP/T001011/1 -- the Quantum Communications Hub and  EP/Z533208/1 -- the Integrated Quantum Networks (IQN) Hub which are both part of the UK national quantum technologies programme. 
It was also supported by the EPSRC new investigator award EP/X039439/1 -- Towards The Quantum Internet: Interconnecting Quantum Networks.

\bmsection{Acknowledgment}
We thank Ruizhi (Mark) Yang and Rui Wang for assistance in the concept phase of this work.
We also recognise John G. Rarity for support throughout the work.
We thank Zoe Davidson for assistance understanding typical vendor operated telecommunications networks.

\bmsection{Disclosures}
The authors declare no conflicts of interest.


\end{backmatter}

\pagebreak




\setcounter{equation}{0}
\setcounter{figure}{0}
\setcounter{table}{0}
\setcounter{section}{0}
\setcounter{page}{1}
\renewcommand{\theequation}{S\arabic{equation}}
\renewcommand{\thefigure}{S\arabic{figure}}
\renewcommand{\thesection}{S\arabic{section}}

\title{Optimising physical parameters of a quantum network based on a loss-jitter trade-off: supplemental document}

\begin{abstract*}
When implementing quantum communications into real world environments, the system have a large bank of parameters which can be optimised to improve performance.
This work pushes towards full optimisation of quantum links, removing assumptions which are typically taken for such systems today.
This supplemental document supports the work of Clark, et al.~\cite{Main}.
\end{abstract*}







\section{dark fibre and coexistence}\label{Sec:DarkFibre}

As quantum communications systems have developed, many producers and vendors have moved from using the telecommunications C-Band ($1530$\,nm to $1565$\,nm) into the telecommunications O-Band ($1260$\,nm to $1360$\,nm).
This has been driven by the notion that classical telecommunications signals are transmitted in the C-Band and the L-band ($1565$\,nm to $1625$\,nm), leaving a $200$\,nm gap between the classical signals and the quantum signals now in the O-band.
This gap is needed to prevent overlap between Raman scattered noise spectrum, which is produced co-propagating with the signals in a wide bandwidth around the wavelength of a bright signal, with the spectrum of the quantum signals~\cite{Bahrani2016OptimalNetworks}.
Here, one single fibre can then be used for both bands with quantum in one and classical in the other, and wavelength multiplexing is then used to separate out the signals\cite{thomas2023designing,mehdizadeh2024quantum}.
Typically referred to as coexistence, this scheme is used to limit to cost of having to use separate optical fibres for the two signals, which would be required if both were in the same band.
When considering long distance terrestrial fibre links, such as cross continent links, this fibre pair cost can significantly increase.
This leads to the thought that coexistence will be the required format for future communication links.

At the same time, classical communications research is expanding from the C+L-Band approach, which is in use by telecommunications industry.
A multi-band communications hero demonstration has simulated record transmission rate of $339.1$\,Tb/s using all 6 of the defined telecommunications bands~\cite{puttnam2025HighTransmission} (the OESCLU-band).
This push to use all optical bands for classical communications is driven by the same cost requirement as coexistence of quantum and classical signals.
However, such demonstrations still rely on custom build components to complete the communications.

The development of optical components has continued across the OESCLU-band, pushing towards scalable and cost effective deployment of OESCLU-band technologies.
This shows that classical telecommunications industry is developing these technologies in line with quantum communications technologies.
A critical technology required is efficient, high-gain, amplifiers across the OESCLU-band.
Here, there are demonstrations in the O-Band~\cite{Dawson:17}, E-Band~\cite{Mikhailov:22}, S-Band~\cite{Aozasa:06}, and L+U-Bands~\cite{mirza2022design}, along with standard Erbium-Doped Fibre Amplifiers for the C+L-Bands~\cite{KeyopsysC, keyopsysL}. 
Such development is also underway for devices such as efficient \gls*{wdm} technology~\cite{Elson24, HAN2025EbandDWDM, Li22ObandDWDM, Wang23CLbandDWDM} and power efficient transmitters~\cite{Kuo18EbandTransmitter}.
Further developments include the use of wavelength conversion to utilise pre-existing optical communication technology, for transmission in the wider bands~\cite{kato2024unleashing, Vitali24LUbandConversion}

This use of the full optical spectrum for classical communications will remove any benefit gained from coexistence, and the requirement for a large spectral gap ($\approx200$\,nm) between the classical and quantum communications is no longer cost effective.
As such, having one full capacity fibre for classical communications and one additional dark fibre for quantum communications is another format for future communication links.
In this case the full optical channel capacity of the dark fibre can also be used for many multiplexed quantum channels in difference bands.

As our work is looking at optimising the selection of quantum signal wavelength and bandwidth, we have chosen not to consider coexistence as it limits the optimisation parameter set.

\section{Simulation Methodology}\label{Sec:Sim}

The simulation tool consists of a mathematical description of the entanglement distribution protocol, calculating the probability of successful transfer of a BBM92~\cite{BBM92} quantum key distribution (QKD) protocol.
Within this description, we can separate the effects on the transmission to three main dependences; the loss mechanisms, the temporal mechanisms, and the noise mechanisms.

To define the overall transmission on the links we determine typical values for; the source heralding efficiency ($\eta_{HE}$), the single photon detector efficiency ($\eta_d$), the optical fibre length to a node ($l_{User}$), the optical fibre attenuation ($\alpha$), the quantum measurement efficiency at each node ($\eta_{User}$), and an additional transmission value related to any instantaneous loss not already considered ($\eta_{I}$).
So when determining the total system transmission between two users, Alice (A) and Bob (B), we recover,
\begin{equation}\label{eqn:transmission}
    \eta_{Total} = \eta_{HE} \times \eta_d \eta_A 10^{\frac{l_{A}\alpha}{10}}  \times \eta_d \eta_B 10^{\frac{l_{B}\alpha}{10}} \times \eta_I.
\end{equation}

For entangled photon pairs, the temporal mechanisms are coupled between the two photons.
As such, many processes that are expected on the transmission of classical light now occur non-locally between the entangled photons.
Specifically, both the relation between the bandwidth and the temporal length, and the chromatic dispersion must now be treated as non-local.
The time-bandwidth effect means that when you temporally measure one of the photons, the arrival time of the second has an uncertainty defined by the time-bandwidth product.
If we take the spectral distribution of the photons to be Gaussian in nature, then the temporal length and the spectral distribution are related by,
\begin{equation}\label{eqn:timebandwidthproduct}
    \tau =  \frac{2 \ln(2)}{\pi} \Delta\nu,
\end{equation}
where $\tau$ is the temporal uncertainty in photon pair simultaneity and $\Delta\nu$ is the frequency bandwidth of the photons.

Optical fibre dispersion is typically dominated by \gls*{cd} ($D(\lambda)$) for standard telecommunications bandwidths, leading to a negligible component of \gls*{pmd} in both cases.
As we consider quantum communications, extremely narrow bandwidth outputs are possible, leading to \gls*{pmd} contribution greater than the \gls*{cd}.
However, in this case the time-bandwidth uncertainty relation means that the temporal length of the photons become the dominant factor.
Here we do not consider the \gls*{pmd} as the contribution remains negligible compare to the other parameters.
The \gls*{cd} is dependent on the wavelength of the signal, the bandwidth of the signal, and the propagation distance, giving a total link \gls*{cd} of,
\begin{equation}\label{eqn:chromaticdispersion}
    \Delta t_{CD} = 
    D(\lambda_{A})l_{A}\frac{\Delta\nu\lambda_{A}^2}{c} + D(\lambda_{B})l_{B}\frac{\Delta\nu\lambda_{B}^2}{c},
\end{equation}
where $\Delta\nu$ is the frequency bandwidth of the photons, $\lambda_{User}$ is the wavelength of the photon going to the User, $l_{User}$ is the link length to the User, $c$ is the speed of light, and $D(\lambda)$ is the \gls*{cd} seen by light of wavelength $\lambda$.

This leads to a total temporal width of the correlation measurement of,
\begin{equation}\label{eqn:temporalwidth}
    \Delta t^2 = 
    \tau^{2} 
    +\Delta t_{CD}^{2}
    + 2\left(J_d^{2}\right) + J_{sync}^{2}
    ,
\end{equation}
where $J_d$ is the measurement jitter of the detectors and $J_{sync}$ is the time synchronisation jitter.
It is useful to define a term for the total measurement jitter, $JT^2 = 2\left(J_d^{2}\right) + J_{sync}^{2}$, which can be substituted for the components to track the total measurement jitter in the system.
This temporal width defines the coincidence window, which is the cross-correlation width used to measure coincidences between the two remote locations.
As this temporal width grows, the coincidence window typically increases proportionally to encapsulate a large proportion of the measured photon pairs.

The noise mechanisms have dominant contributions from; additional photons transmitted through the optical fibre that also reach the single photon detectors, loss of one photon in the entangled pair, and the inherent dark count rates for the single photon detectors.
As the measurements in this type of system are correlations, this increases the number of accidental coincidences that are recovered within the coincidence window, reducing the \gls*{car}.
When performing BBM92 \gls*{qkd}, this directly increases the \gls*{qber} reducing the \gls*{skr}.

The Simulation tool takes in all of the above inputs, along with the terms for the defined coincidence window and the entangled photon pair generation rate.
This tool was tested against experimental date from Joshi et. al.~\cite{joshi2020trusted}, Wang et. al.~\cite{wang2022dynamic,wang2023field,wang2023optimum}, and Clark et. al.~\cite{clark2023entanglement,clark2024quantum} showing the accuracy of the tool for simulating the links.

By using a scalar function minimisation, we can compute the values of coincidence window and entangled photon pair generation rate that lead to the highest \gls*{skr}.
This output set of parameters are then taken to be the best-case performance for a link with the set of specific inputs.
We then repeat this minimisation for all combinations of fibre length, total measurement jitter, wavelength, bandwidth, chromatic dispersion, and attenuation.

\section{Constructing the optimisation mosaic}\label{Sec:Mosaic}

Using the Simulator tool, discussed in Section \ref{Sec:Sim}, we construct an optimisation mosaic showing the optimal selection for a given combination of input parameters.
An optimisation mosaic is a heat-map defining regions where one parameter is the optimal in a discrete set of inputs, instead of representing a direct output value from the simulator.
Below is the process through which the comparison between different wavelengths, bandwidths, and fibres are completed when only a single quantum communications channel is transmitted through the optical fibre.

\begin{figure}[!b]
    \centering
    \includegraphics[width=13cm]{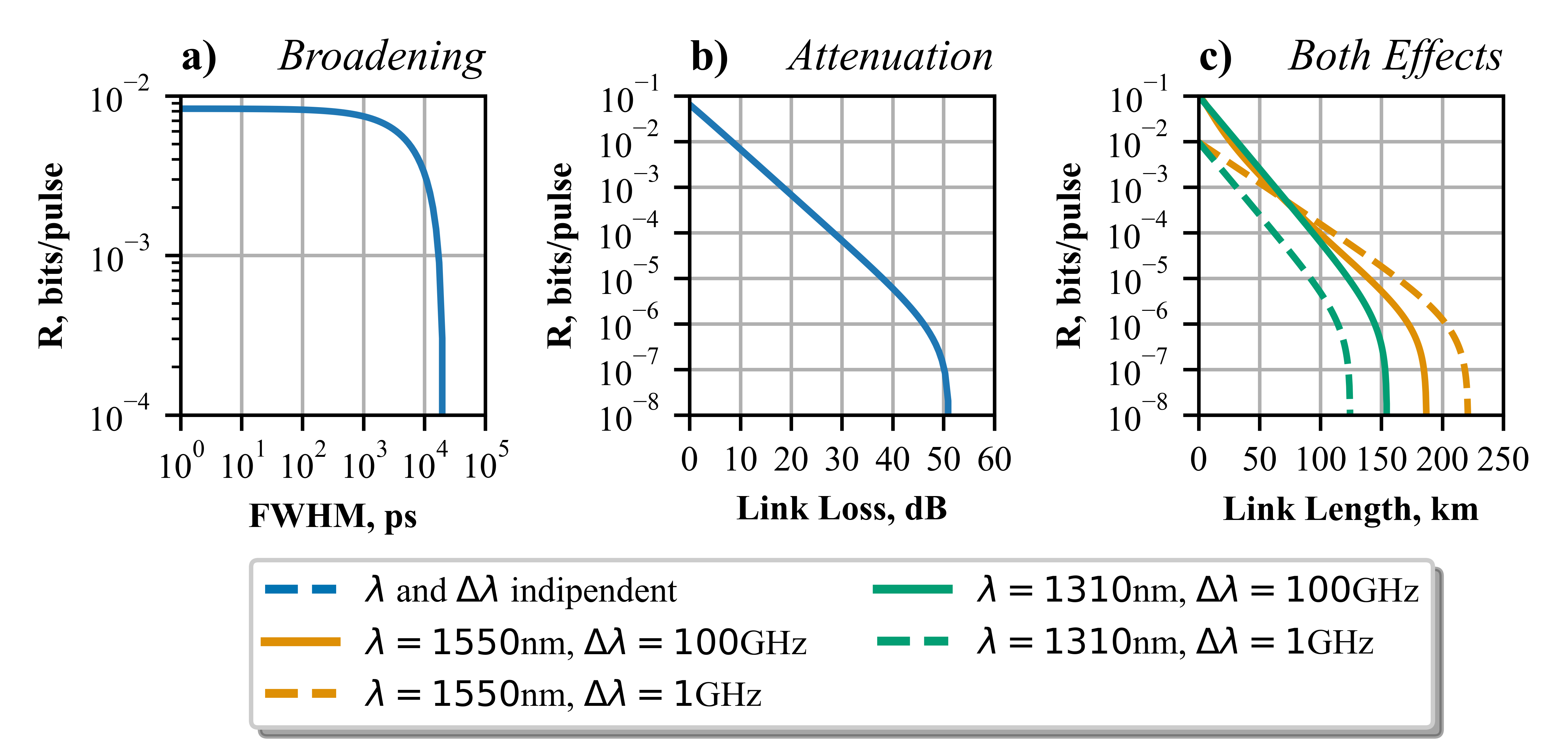}
    \caption{
    The simulation results based on combinations of inputs, showing the rate of correlations seen per pulse, here defined to be a discrete time width.
    \textbf{a)} shows the effect of increasing the \acrfull*{fwhm}.
    \textbf{b)} shows the effect of increasing the attenuation of the link.
    \textbf{c)} shows the effect of a single mode fibre~\cite{CorningUltra} link combining both the increase of the \gls*{fwhm} and attenuation.
    Here the fibre has chromatic dispersion of \SI{18}{\pico\second\per\nano\meter\per\kilo\meter} and attenuation of \SI{0.18}{\decibel\per\kilo\meter}.
    }
    \label{fig:BasicSims}
\end{figure}

To construct such a data set, we must first understand how the simulator outputs relate to the inputs.
Fig.~\ref{fig:BasicSims} shows the how the \gls*{skr} value recovered from the simulator depends on one parameter.
In Fig.~\ref{fig:BasicSims} \textbf{a)} the relationship between the \gls*{fwhm} of a correlation histogram affects the rate of correlations.
We can see, for a fixed noise scenario, that at a certain point the correlation rate significantly reduces to a noise floor and the rate goes to zero. 
If instead we fix the \gls*{fwhm} and instead change the link loss, as shown in \ref{fig:BasicSims} \textbf{b)}, we can see a more linear relationship between the rate and the link loss, again until the correlation rate hits a noise floor and goes to zero.

The amount the \gls*{fwhm} and link loss dependents on the fibre length depends on both the wavelength and the bandwidth of the signal.
Fig.~\ref{fig:BasicSims} \textbf{c)} shows the case for standard telecommunications \gls*{smf}~\cite{CorningUltra}, for both $1550$\,nm and $1310$\,nm.
Both wavelengths are displayed for both the $100$\,GHz and $1$\,GHz cases.
This figure shows that when the bandwidth is larger and the fibre length is low, the correlation rate is higher than with a lower bandwidth.
This comes from the time bandwidth relation, so a narrower bandwidth leads to a larger temporal width and so a larger \gls*{fwhm}, decreasing the correlation rate.
When \gls*{cd} is negligible, at $1310$\,nm, we can see that the only further dependence is on the attenuation so both reduce the correlational rate proportionally.
When \gls*{cd} increases the \gls*{fwhm} over fibre length, we get an additional dependence on the bandwidth.
This is due to the dependence on both fibre length and bandwidth in \gls*{cd}.
For $1550$\,nm we then see the correlation rate reduce faster for the larger bandwidth case, and the link fails at a short distance, even with a higher correlation rate at $0$\,km.

\begin{figure}[!b]
    \centering
    \includegraphics[width=13cm]{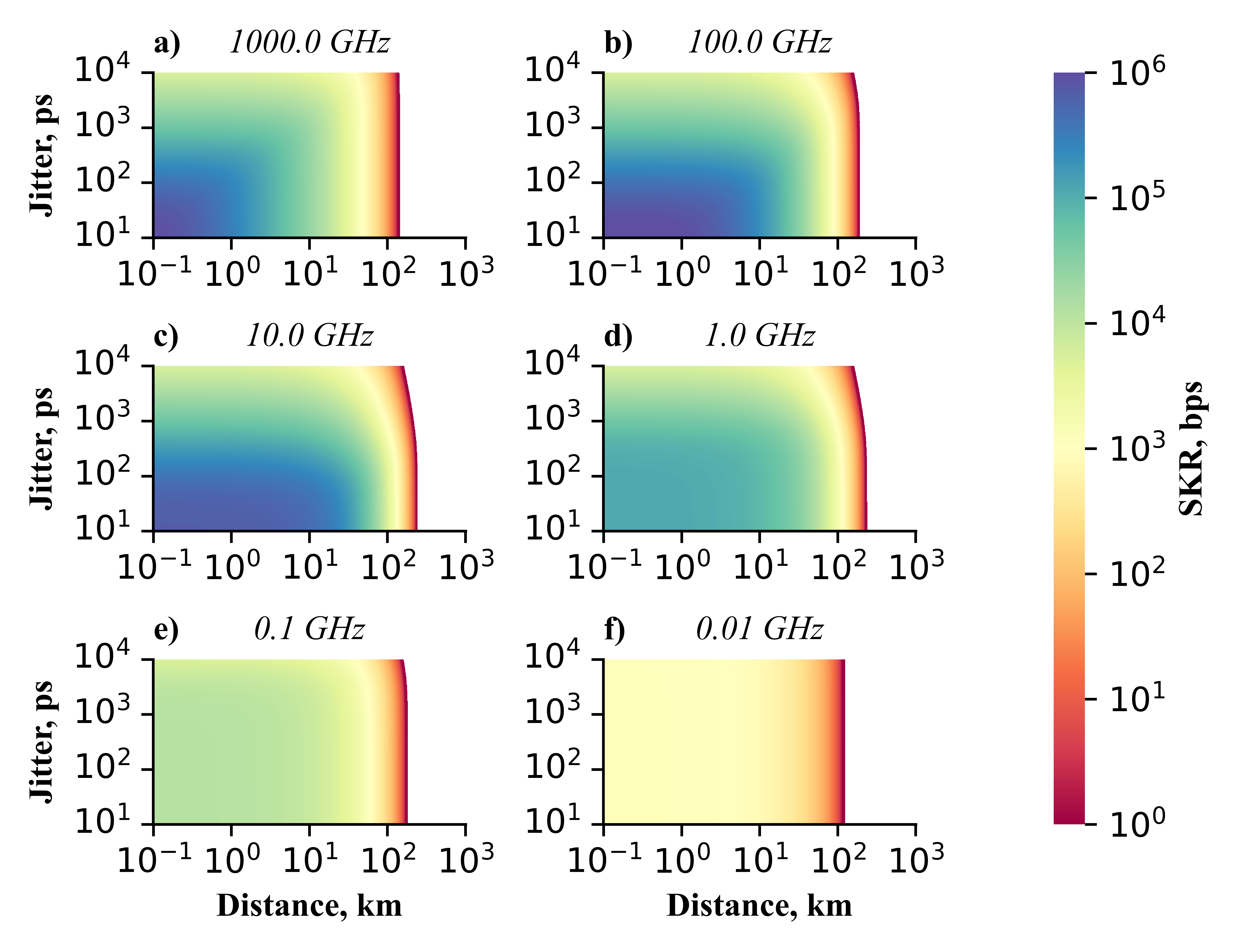}
    \caption{
    The \acrfull*{skr} of a BBM-92 \gls*{qkd}~\cite{BBM92}, with chromatic dispersion of \SI{18}{\pico\second\per\nano\meter\per\kilo\meter} and attenuation of \SI{0.18}{\decibel\per\kilo\meter} 
    as a function of the total timing jitter and fibre length, for a selection of photon spectral bandwidths.
    The bandwidth represented in each subplot is given in the subplot title.
    }
    \label{fig:SKRHeatmaps}
\end{figure}

As the fibre length changes both the \gls*{cd} and the attenuation, we can now combine the the fibre length effects with the total measurement jitter ($JT$), as discussed in Section \ref{Sec:Sim}, and produce heat-maps for each combination of wavelength and bandwidths.
Fig.~\ref{fig:SKRHeatmaps} shows example results for a \SI{1550}{\nano\meter} light.
These heat-maps show the \gls*{skr} of the links, but direct comparison between heat-maps is difficult.
From this, we condense the information into a single heat-map by comparing all the heat-maps for all combinations of jitter and fibre distance.
The optimisation mosaic now shows the bandwidth that returned the highest \gls*{skr}.
This is done for each of the fibre and wavelength cases, producing the mosaics shown in Fig.~2 and Fig.~3.
This allows for comparison between the optimal bandwidth for three cases;
\gls*{smf}~\cite{CorningUltra} with \SI{1550}{\nano\meter},
\gls*{smf}~\cite{CorningUltra} with \SI{1310}{\nano\meter},
and \gls*{nzdsf}~\cite{CorningLEAF} with \SI{1550}{\nano\meter}.
However, this now gives no information to compare the different fibre and wavelength combinations.

To directly compare the different cases, we add another round of comparison.
By selecting the \gls*{skr} for all bandwidths in all combinations of optical fibre and wavelength we recover a single mosaic, giving the true optimal set of parameters for a link with a given measurement jitter and fibre length, as shown in Fig.~4 and Fig.~5.

\newpage
\begin{landscape}
\section{Detailed Data Breakdown}\label{Sec:Tables}

\begin{table}[ht]
\centering
\begin{tabular}{|l|l|l|lllllllllllll|}
\hline
 &
   &
   &
  \multicolumn{13}{l|}{Bandwidth, GHz} \\
\multirow{-2}{*}{Jitter} &
  \multirow{-2}{*}{Fibre} &
  \multirow{-2}{*}{Wavelength} &
  \multicolumn{1}{l|}{1000} &
  \multicolumn{1}{l|}{500} &
  \multicolumn{1}{l|}{200} &
  \multicolumn{1}{l|}{100} &
  \multicolumn{1}{l|}{50} &
  \multicolumn{1}{l|}{20} &
  \multicolumn{1}{l|}{10} &
  \multicolumn{1}{l|}{5} &
  \multicolumn{1}{l|}{2} &
  \multicolumn{1}{l|}{1} &
  \multicolumn{1}{l|}{0.5} &
  \multicolumn{1}{l|}{0.2} &
  0.1 \\ \hline
 &
  \cellcolor[HTML]{E8E8E8}SMF &
  \cellcolor[HTML]{E8E8E8}1550nm &
  \multicolumn{1}{l|}{\cellcolor[HTML]{E8E8E8}0.013} &
  \multicolumn{1}{l|}{\cellcolor[HTML]{E8E8E8}0.03} &
  \multicolumn{1}{l|}{\cellcolor[HTML]{E8E8E8}0.08} &
  \multicolumn{1}{l|}{\cellcolor[HTML]{E8E8E8}0.163} &
  \multicolumn{1}{l|}{\cellcolor[HTML]{E8E8E8}0.319} &
  \multicolumn{1}{l|}{\cellcolor[HTML]{E8E8E8}0.668} &
  \multicolumn{1}{l|}{\cellcolor[HTML]{E8E8E8}0.923} &
  \multicolumn{1}{l|}{\cellcolor[HTML]{E8E8E8}1} &
  \multicolumn{1}{l|}{\cellcolor[HTML]{E8E8E8}0.78} &
  \multicolumn{1}{l|}{\cellcolor[HTML]{E8E8E8}0.489} &
  \multicolumn{1}{l|}{\cellcolor[HTML]{E8E8E8}0.264} &
  \multicolumn{1}{l|}{\cellcolor[HTML]{E8E8E8}0.106} &
  \cellcolor[HTML]{E8E8E8}0.051 \\ \cline{2-16} 
 &
  SMF &
  1310nm &
  \multicolumn{1}{l|}{0.041} &
  \multicolumn{1}{l|}{0.041} &
  \multicolumn{1}{l|}{0.041} &
  \multicolumn{1}{l|}{0.041} &
  \multicolumn{1}{l|}{0.041} &
  \multicolumn{1}{l|}{0.041} &
  \multicolumn{1}{l|}{0.039} &
  \multicolumn{1}{l|}{0.035} &
  \multicolumn{1}{l|}{0.021} &
  \multicolumn{1}{l|}{0.011} &
  \multicolumn{1}{l|}{0.004} &
  \multicolumn{1}{l|}{0} &
  0 \\ \cline{2-16} 
\multirow{-3}{*}{200} &
  \cellcolor[HTML]{D7F4F5}NZDSF &
  \cellcolor[HTML]{D7F4F5}1550nm &
  \multicolumn{1}{l|}{\cellcolor[HTML]{D7F4F5}0.056} &
  \multicolumn{1}{l|}{\cellcolor[HTML]{D7F4F5}0.116} &
  \multicolumn{1}{l|}{\cellcolor[HTML]{D7F4F5}0.281} &
  \multicolumn{1}{l|}{\cellcolor[HTML]{D7F4F5}0.494} &
  \multicolumn{1}{l|}{\cellcolor[HTML]{D7F4F5}0.717} &
  \multicolumn{1}{l|}{\cellcolor[HTML]{D7F4F5}0.861} &
  \multicolumn{1}{l|}{\cellcolor[HTML]{D7F4F5}0.876} &
  \multicolumn{1}{l|}{\cellcolor[HTML]{D7F4F5}0.831} &
  \multicolumn{1}{l|}{\cellcolor[HTML]{D7F4F5}0.621} &
  \multicolumn{1}{l|}{\cellcolor[HTML]{D7F4F5}0.388} &
  \multicolumn{1}{l|}{\cellcolor[HTML]{D7F4F5}0.209} &
  \multicolumn{1}{l|}{\cellcolor[HTML]{D7F4F5}0.083} &
  \cellcolor[HTML]{D7F4F5}0.04 \\ \hline
 &
  \cellcolor[HTML]{E8E8E8}SMF &
  \cellcolor[HTML]{E8E8E8}1550nm &
  \multicolumn{1}{l|}{\cellcolor[HTML]{E8E8E8}0.006} &
  \multicolumn{1}{l|}{\cellcolor[HTML]{E8E8E8}0.014} &
  \multicolumn{1}{l|}{\cellcolor[HTML]{E8E8E8}0.037} &
  \multicolumn{1}{l|}{\cellcolor[HTML]{E8E8E8}0.076} &
  \multicolumn{1}{l|}{\cellcolor[HTML]{E8E8E8}0.152} &
  \multicolumn{1}{l|}{\cellcolor[HTML]{E8E8E8}0.367} &
  \multicolumn{1}{l|}{\cellcolor[HTML]{E8E8E8}0.65} &
  \multicolumn{1}{l|}{\cellcolor[HTML]{E8E8E8}0.809} &
  \multicolumn{1}{l|}{\cellcolor[HTML]{E8E8E8}0.465} &
  \multicolumn{1}{l|}{\cellcolor[HTML]{E8E8E8}0.246} &
  \multicolumn{1}{l|}{\cellcolor[HTML]{E8E8E8}0.125} &
  \multicolumn{1}{l|}{\cellcolor[HTML]{E8E8E8}0.049} &
  \cellcolor[HTML]{E8E8E8}0.024 \\ \cline{2-16} 
 &
  SMF &
  1310nm &
  \multicolumn{1}{l|}{0.067} &
  \multicolumn{1}{l|}{0.067} &
  \multicolumn{1}{l|}{0.067} &
  \multicolumn{1}{l|}{0.066} &
  \multicolumn{1}{l|}{0.065} &
  \multicolumn{1}{l|}{0.059} &
  \multicolumn{1}{l|}{0.046} &
  \multicolumn{1}{l|}{0.029} &
  \multicolumn{1}{l|}{0.012} &
  \multicolumn{1}{l|}{0.005} &
  \multicolumn{1}{l|}{0.002} &
  \multicolumn{1}{l|}{0} &
  0 \\ \cline{2-16} 
\multirow{-3}{*}{50} &
  \cellcolor[HTML]{D7F4F5}NZDSF &
  \cellcolor[HTML]{D7F4F5}1550nm &
  \multicolumn{1}{l|}{\cellcolor[HTML]{D7F4F5}0.026} &
  \multicolumn{1}{l|}{\cellcolor[HTML]{D7F4F5}0.054} &
  \multicolumn{1}{l|}{\cellcolor[HTML]{D7F4F5}0.136} &
  \multicolumn{1}{l|}{\cellcolor[HTML]{D7F4F5}0.264} &
  \multicolumn{1}{l|}{\cellcolor[HTML]{D7F4F5}0.49} &
  \multicolumn{1}{l|}{\cellcolor[HTML]{D7F4F5}0.899} &
  \multicolumn{1}{l|}{\cellcolor[HTML]{D7F4F5}1} &
  \multicolumn{1}{l|}{\cellcolor[HTML]{D7F4F5}0.76} &
  \multicolumn{1}{l|}{\cellcolor[HTML]{D7F4F5}0.372} &
  \multicolumn{1}{l|}{\cellcolor[HTML]{D7F4F5}0.195} &
  \multicolumn{1}{l|}{\cellcolor[HTML]{D7F4F5}0.099} &
  \multicolumn{1}{l|}{\cellcolor[HTML]{D7F4F5}0.039} &
  \cellcolor[HTML]{D7F4F5}0.018 \\ \hline
 &
  \cellcolor[HTML]{E8E8E8}SMF &
  \cellcolor[HTML]{E8E8E8}1550nm &
  \multicolumn{1}{l|}{\cellcolor[HTML]{E8E8E8}0.005} &
  \multicolumn{1}{l|}{\cellcolor[HTML]{E8E8E8}0.011} &
  \multicolumn{1}{l|}{\cellcolor[HTML]{E8E8E8}0.029} &
  \multicolumn{1}{l|}{\cellcolor[HTML]{E8E8E8}0.06} &
  \multicolumn{1}{l|}{\cellcolor[HTML]{E8E8E8}0.12} &
  \multicolumn{1}{l|}{\cellcolor[HTML]{E8E8E8}0.293} &
  \multicolumn{1}{l|}{\cellcolor[HTML]{E8E8E8}0.536} &
  \multicolumn{1}{l|}{\cellcolor[HTML]{E8E8E8}0.688} &
  \multicolumn{1}{l|}{\cellcolor[HTML]{E8E8E8}0.375} &
  \multicolumn{1}{l|}{\cellcolor[HTML]{E8E8E8}0.195} &
  \multicolumn{1}{l|}{\cellcolor[HTML]{E8E8E8}0.098} &
  \multicolumn{1}{l|}{\cellcolor[HTML]{E8E8E8}0.039} &
  \cellcolor[HTML]{E8E8E8}0.019 \\ \cline{2-16} 
 &
  SMF &
  1310nm &
  \multicolumn{1}{l|}{0.165} &
  \multicolumn{1}{l|}{0.164} &
  \multicolumn{1}{l|}{0.158} &
  \multicolumn{1}{l|}{0.145} &
  \multicolumn{1}{l|}{0.118} &
  \multicolumn{1}{l|}{0.074} &
  \multicolumn{1}{l|}{0.044} &
  \multicolumn{1}{l|}{0.024} &
  \multicolumn{1}{l|}{0.009} &
  \multicolumn{1}{l|}{0.004} &
  \multicolumn{1}{l|}{0.001} &
  \multicolumn{1}{l|}{0} &
  0 \\ \cline{2-16} 
\multirow{-3}{*}{5} &
  \cellcolor[HTML]{D7F4F5}NZDSF &
  \cellcolor[HTML]{D7F4F5}1550nm &
  \multicolumn{1}{l|}{\cellcolor[HTML]{D7F4F5}0.02} &
  \multicolumn{1}{l|}{\cellcolor[HTML]{D7F4F5}0.042} &
  \multicolumn{1}{l|}{\cellcolor[HTML]{D7F4F5}0.107} &
  \multicolumn{1}{l|}{\cellcolor[HTML]{D7F4F5}0.211} &
  \multicolumn{1}{l|}{\cellcolor[HTML]{D7F4F5}0.403} &
  \multicolumn{1}{l|}{\cellcolor[HTML]{D7F4F5}0.844} &
  \multicolumn{1}{l|}{\cellcolor[HTML]{D7F4F5}1} &
  \multicolumn{1}{l|}{\cellcolor[HTML]{D7F4F5}0.671} &
  \multicolumn{1}{l|}{\cellcolor[HTML]{D7F4F5}0.3} &
  \multicolumn{1}{l|}{\cellcolor[HTML]{D7F4F5}0.155} &
  \multicolumn{1}{l|}{\cellcolor[HTML]{D7F4F5}0.078} &
  \multicolumn{1}{l|}{\cellcolor[HTML]{D7F4F5}0.03} &
  \cellcolor[HTML]{D7F4F5}0.014 \\ \hline
\end{tabular}
    \caption{
    The normalised SKR for a distance of \SI{100}{\kilo\meter} with only a single quantum channel. 
    Here the SKR is normalised to the maximum value seen in the combination of distance and jitter.
    }
    \label{tab:3.1.1-summary-100km}
\end{table}

\newpage
\begin{table}[ht]
\centering
\begin{tabular}{|l|l|l|lllllllllllll|}
\hline
 &
   &
   &
  \multicolumn{13}{l|}{Bandwidth, GHz} \\
\multirow{-2}{*}{Jitter} &
  \multirow{-2}{*}{Fibre} &
  \multirow{-2}{*}{Wavelength} &
  \multicolumn{1}{l|}{1000} &
  \multicolumn{1}{l|}{500} &
  \multicolumn{1}{l|}{200} &
  \multicolumn{1}{l|}{100} &
  \multicolumn{1}{l|}{50} &
  \multicolumn{1}{l|}{20} &
  \multicolumn{1}{l|}{10} &
  \multicolumn{1}{l|}{5} &
  \multicolumn{1}{l|}{2} &
  \multicolumn{1}{l|}{1} &
  \multicolumn{1}{l|}{0.5} &
  \multicolumn{1}{l|}{0.2} &
  0.1 \\ \hline
 &
  \cellcolor[HTML]{E8E8E8}SMF &
  \cellcolor[HTML]{E8E8E8}1550nm &
  \multicolumn{1}{l|}{\cellcolor[HTML]{E8E8E8}0.155} &
  \multicolumn{1}{l|}{\cellcolor[HTML]{E8E8E8}0.296} &
  \multicolumn{1}{l|}{\cellcolor[HTML]{E8E8E8}0.609} &
  \multicolumn{1}{l|}{\cellcolor[HTML]{E8E8E8}0.841} &
  \multicolumn{1}{l|}{\cellcolor[HTML]{E8E8E8}0.96} &
  \multicolumn{1}{l|}{\cellcolor[HTML]{E8E8E8}1} &
  \multicolumn{1}{l|}{\cellcolor[HTML]{E8E8E8}0.991} &
  \multicolumn{1}{l|}{\cellcolor[HTML]{E8E8E8}0.937} &
  \multicolumn{1}{l|}{\cellcolor[HTML]{E8E8E8}0.708} &
  \multicolumn{1}{l|}{\cellcolor[HTML]{E8E8E8}0.45} &
  \multicolumn{1}{l|}{\cellcolor[HTML]{E8E8E8}0.247} &
  \multicolumn{1}{l|}{\cellcolor[HTML]{E8E8E8}0.102} &
  \cellcolor[HTML]{E8E8E8}0.052 \\ \cline{2-16} 
 &
  SMF &
  1310nm &
  \multicolumn{1}{l|}{0.745} &
  \multicolumn{1}{l|}{0.745} &
  \multicolumn{1}{l|}{0.745} &
  \multicolumn{1}{l|}{0.744} &
  \multicolumn{1}{l|}{0.743} &
  \multicolumn{1}{l|}{0.737} &
  \multicolumn{1}{l|}{0.715} &
  \multicolumn{1}{l|}{0.644} &
  \multicolumn{1}{l|}{0.425} &
  \multicolumn{1}{l|}{0.247} &
  \multicolumn{1}{l|}{0.13} &
  \multicolumn{1}{l|}{0.053} &
  0.027 \\ \cline{2-16} 
\multirow{-3}{*}{200} &
  \cellcolor[HTML]{D7F4F5}NZDSF &
  \cellcolor[HTML]{D7F4F5}1550nm &
  \multicolumn{1}{l|}{\cellcolor[HTML]{D7F4F5}0.556} &
  \multicolumn{1}{l|}{\cellcolor[HTML]{D7F4F5}0.795} &
  \multicolumn{1}{l|}{\cellcolor[HTML]{D7F4F5}0.95} &
  \multicolumn{1}{l|}{\cellcolor[HTML]{D7F4F5}0.981} &
  \multicolumn{1}{l|}{\cellcolor[HTML]{D7F4F5}0.989} &
  \multicolumn{1}{l|}{\cellcolor[HTML]{D7F4F5}0.987} &
  \multicolumn{1}{l|}{\cellcolor[HTML]{D7F4F5}0.972} &
  \multicolumn{1}{l|}{\cellcolor[HTML]{D7F4F5}0.917} &
  \multicolumn{1}{l|}{\cellcolor[HTML]{D7F4F5}0.693} &
  \multicolumn{1}{l|}{\cellcolor[HTML]{D7F4F5}0.44} &
  \multicolumn{1}{l|}{\cellcolor[HTML]{D7F4F5}0.241} &
  \multicolumn{1}{l|}{\cellcolor[HTML]{D7F4F5}0.1} &
  \cellcolor[HTML]{D7F4F5}0.05 \\ \hline
 &
  \cellcolor[HTML]{E8E8E8}SMF &
  \cellcolor[HTML]{E8E8E8}1550nm &
  \multicolumn{1}{l|}{\cellcolor[HTML]{E8E8E8}0.056} &
  \multicolumn{1}{l|}{\cellcolor[HTML]{E8E8E8}0.11} &
  \multicolumn{1}{l|}{\cellcolor[HTML]{E8E8E8}0.258} &
  \multicolumn{1}{l|}{\cellcolor[HTML]{E8E8E8}0.459} &
  \multicolumn{1}{l|}{\cellcolor[HTML]{E8E8E8}0.714} &
  \multicolumn{1}{l|}{\cellcolor[HTML]{E8E8E8}0.923} &
  \multicolumn{1}{l|}{\cellcolor[HTML]{E8E8E8}0.862} &
  \multicolumn{1}{l|}{\cellcolor[HTML]{E8E8E8}0.64} &
  \multicolumn{1}{l|}{\cellcolor[HTML]{E8E8E8}0.326} &
  \multicolumn{1}{l|}{\cellcolor[HTML]{E8E8E8}0.175} &
  \multicolumn{1}{l|}{\cellcolor[HTML]{E8E8E8}0.091} &
  \multicolumn{1}{l|}{\cellcolor[HTML]{E8E8E8}0.037} &
  \cellcolor[HTML]{E8E8E8}0.019 \\ \cline{2-16} 
 &
  SMF &
  1310nm &
  \multicolumn{1}{l|}{0.801} &
  \multicolumn{1}{l|}{0.801} &
  \multicolumn{1}{l|}{0.8} &
  \multicolumn{1}{l|}{0.797} &
  \multicolumn{1}{l|}{0.785} &
  \multicolumn{1}{l|}{0.718} &
  \multicolumn{1}{l|}{0.576} &
  \multicolumn{1}{l|}{0.38} &
  \multicolumn{1}{l|}{0.177} &
  \multicolumn{1}{l|}{0.093} &
  \multicolumn{1}{l|}{0.047} &
  \multicolumn{1}{l|}{0.019} &
  0.01 \\ \cline{2-16} 
\multirow{-3}{*}{50} &
  \cellcolor[HTML]{D7F4F5}NZDSF &
  \cellcolor[HTML]{D7F4F5}1550nm &
  \multicolumn{1}{l|}{\cellcolor[HTML]{D7F4F5}0.23} &
  \multicolumn{1}{l|}{\cellcolor[HTML]{D7F4F5}0.414} &
  \multicolumn{1}{l|}{\cellcolor[HTML]{D7F4F5}0.743} &
  \multicolumn{1}{l|}{\cellcolor[HTML]{D7F4F5}0.928} &
  \multicolumn{1}{l|}{\cellcolor[HTML]{D7F4F5}1} &
  \multicolumn{1}{l|}{\cellcolor[HTML]{D7F4F5}0.979} &
  \multicolumn{1}{l|}{\cellcolor[HTML]{D7F4F5}0.858} &
  \multicolumn{1}{l|}{\cellcolor[HTML]{D7F4F5}0.628} &
  \multicolumn{1}{l|}{\cellcolor[HTML]{D7F4F5}0.319} &
  \multicolumn{1}{l|}{\cellcolor[HTML]{D7F4F5}0.171} &
  \multicolumn{1}{l|}{\cellcolor[HTML]{D7F4F5}0.089} &
  \multicolumn{1}{l|}{\cellcolor[HTML]{D7F4F5}0.036} &
  \cellcolor[HTML]{D7F4F5}0.018 \\ \hline
 &
  \cellcolor[HTML]{E8E8E8}SMF &
  \cellcolor[HTML]{E8E8E8}1550nm &
  \multicolumn{1}{l|}{\cellcolor[HTML]{E8E8E8}0.028} &
  \multicolumn{1}{l|}{\cellcolor[HTML]{E8E8E8}0.054} &
  \multicolumn{1}{l|}{\cellcolor[HTML]{E8E8E8}0.129} &
  \multicolumn{1}{l|}{\cellcolor[HTML]{E8E8E8}0.237} &
  \multicolumn{1}{l|}{\cellcolor[HTML]{E8E8E8}0.405} &
  \multicolumn{1}{l|}{\cellcolor[HTML]{E8E8E8}0.631} &
  \multicolumn{1}{l|}{\cellcolor[HTML]{E8E8E8}0.546} &
  \multicolumn{1}{l|}{\cellcolor[HTML]{E8E8E8}0.35} &
  \multicolumn{1}{l|}{\cellcolor[HTML]{E8E8E8}0.164} &
  \multicolumn{1}{l|}{\cellcolor[HTML]{E8E8E8}0.087} &
  \multicolumn{1}{l|}{\cellcolor[HTML]{E8E8E8}0.045} &
  \multicolumn{1}{l|}{\cellcolor[HTML]{E8E8E8}0.018} &
  \cellcolor[HTML]{E8E8E8}0.009 \\ \cline{2-16} 
 &
  SMF &
  1310nm &
  \multicolumn{1}{l|}{1} &
  \multicolumn{1}{l|}{0.996} &
  \multicolumn{1}{l|}{0.97} &
  \multicolumn{1}{l|}{0.904} &
  \multicolumn{1}{l|}{0.772} &
  \multicolumn{1}{l|}{0.526} &
  \multicolumn{1}{l|}{0.34} &
  \multicolumn{1}{l|}{0.199} &
  \multicolumn{1}{l|}{0.088} &
  \multicolumn{1}{l|}{0.046} &
  \multicolumn{1}{l|}{0.023} &
  \multicolumn{1}{l|}{0.009} &
  0.005 \\ \cline{2-16} 
\multirow{-3}{*}{5} &
  \cellcolor[HTML]{D7F4F5}NZDSF &
  \cellcolor[HTML]{D7F4F5}1550nm &
  \multicolumn{1}{l|}{\cellcolor[HTML]{D7F4F5}0.114} &
  \multicolumn{1}{l|}{\cellcolor[HTML]{D7F4F5}0.212} &
  \multicolumn{1}{l|}{\cellcolor[HTML]{D7F4F5}0.433} &
  \multicolumn{1}{l|}{\cellcolor[HTML]{D7F4F5}0.659} &
  \multicolumn{1}{l|}{\cellcolor[HTML]{D7F4F5}0.853} &
  \multicolumn{1}{l|}{\cellcolor[HTML]{D7F4F5}0.777} &
  \multicolumn{1}{l|}{\cellcolor[HTML]{D7F4F5}0.552} &
  \multicolumn{1}{l|}{\cellcolor[HTML]{D7F4F5}0.344} &
  \multicolumn{1}{l|}{\cellcolor[HTML]{D7F4F5}0.16} &
  \multicolumn{1}{l|}{\cellcolor[HTML]{D7F4F5}0.085} &
  \multicolumn{1}{l|}{\cellcolor[HTML]{D7F4F5}0.044} &
  \multicolumn{1}{l|}{\cellcolor[HTML]{D7F4F5}0.018} &
  \cellcolor[HTML]{D7F4F5}0.009 \\ \hline
\end{tabular}
    \caption{
    The normalised SKR for a distance of \SI{10}{\kilo\meter} with only a single quantum channel. 
    Here the SKR is normalised to the maximum value seen in the combination of distance and jitter.
    }
    \label{tab:3.1.1-summary-10km}
\end{table}

\begin{table}[ht]
\centering
\begin{tabular}{|l|l|l|lllllllllllll|}
\hline
 &
   &
   &
  \multicolumn{13}{l|}{Bandwidth, GHz} \\
\multirow{-2}{*}{Jitter} &
  \multirow{-2}{*}{Fibre} &
  \multirow{-2}{*}{Wavelength} &
  \multicolumn{1}{l|}{100} &
  \multicolumn{1}{l|}{50} &
  \multicolumn{1}{l|}{20} &
  \multicolumn{1}{l|}{10} &
  \multicolumn{1}{l|}{5} &
  \multicolumn{1}{l|}{2} &
  \multicolumn{1}{l|}{1} &
  \multicolumn{1}{l|}{0.5} &
  \multicolumn{1}{l|}{0.2} &
  \multicolumn{1}{l|}{0.1} &
  \multicolumn{1}{l|}{0.05} &
  \multicolumn{1}{l|}{0.02} &
  0.01 \\ \hline
 &
  \cellcolor[HTML]{E8E8E8}SMF &
  \cellcolor[HTML]{E8E8E8}1550nm &
  \multicolumn{1}{l|}{\cellcolor[HTML]{E8E8E8}0.003} &
  \multicolumn{1}{l|}{\cellcolor[HTML]{E8E8E8}0.012} &
  \multicolumn{1}{l|}{\cellcolor[HTML]{E8E8E8}0.063} &
  \multicolumn{1}{l|}{\cellcolor[HTML]{E8E8E8}0.174} &
  \multicolumn{1}{l|}{\cellcolor[HTML]{E8E8E8}0.377} &
  \multicolumn{1}{l|}{\cellcolor[HTML]{E8E8E8}0.736} &
  \multicolumn{1}{l|}{\cellcolor[HTML]{E8E8E8}0.923} &
  \multicolumn{1}{l|}{\cellcolor[HTML]{E8E8E8}0.995} &
  \multicolumn{1}{l|}{\cellcolor[HTML]{E8E8E8}1} &
  \multicolumn{1}{l|}{\cellcolor[HTML]{E8E8E8}0.963} &
  \multicolumn{1}{l|}{\cellcolor[HTML]{E8E8E8}0.881} &
  \multicolumn{1}{l|}{\cellcolor[HTML]{E8E8E8}0.662} &
  \cellcolor[HTML]{E8E8E8}0.381 \\ \cline{2-16} 
 &
  SMF &
  1310nm &
  \multicolumn{1}{l|}{0.001} &
  \multicolumn{1}{l|}{0.002} &
  \multicolumn{1}{l|}{0.004} &
  \multicolumn{1}{l|}{0.007} &
  \multicolumn{1}{l|}{0.013} &
  \multicolumn{1}{l|}{0.02} &
  \multicolumn{1}{l|}{0.02} &
  \multicolumn{1}{l|}{0.015} &
  \multicolumn{1}{l|}{0.004} &
  \multicolumn{1}{l|}{0} &
  \multicolumn{1}{l|}{0} &
  \multicolumn{1}{l|}{0} &
  0 \\ \cline{2-16} 
\multirow{-3}{*}{200} &
  \cellcolor[HTML]{D7F4F5}NZDSF &
  \cellcolor[HTML]{D7F4F5}1550nm &
  \multicolumn{1}{l|}{\cellcolor[HTML]{D7F4F5}0.009} &
  \multicolumn{1}{l|}{\cellcolor[HTML]{D7F4F5}0.027} &
  \multicolumn{1}{l|}{\cellcolor[HTML]{D7F4F5}0.081} &
  \multicolumn{1}{l|}{\cellcolor[HTML]{D7F4F5}0.165} &
  \multicolumn{1}{l|}{\cellcolor[HTML]{D7F4F5}0.313} &
  \multicolumn{1}{l|}{\cellcolor[HTML]{D7F4F5}0.586} &
  \multicolumn{1}{l|}{\cellcolor[HTML]{D7F4F5}0.731} &
  \multicolumn{1}{l|}{\cellcolor[HTML]{D7F4F5}0.787} &
  \multicolumn{1}{l|}{\cellcolor[HTML]{D7F4F5}0.785} &
  \multicolumn{1}{l|}{\cellcolor[HTML]{D7F4F5}0.747} &
  \multicolumn{1}{l|}{\cellcolor[HTML]{D7F4F5}0.668} &
  \multicolumn{1}{l|}{\cellcolor[HTML]{D7F4F5}0.462} &
  \cellcolor[HTML]{D7F4F5}0.219 \\ \hline
 &
  \cellcolor[HTML]{E8E8E8}SMF &
  \cellcolor[HTML]{E8E8E8}1550nm &
  \multicolumn{1}{l|}{\cellcolor[HTML]{E8E8E8}0.003} &
  \multicolumn{1}{l|}{\cellcolor[HTML]{E8E8E8}0.012} &
  \multicolumn{1}{l|}{\cellcolor[HTML]{E8E8E8}0.073} &
  \multicolumn{1}{l|}{\cellcolor[HTML]{E8E8E8}0.26} &
  \multicolumn{1}{l|}{\cellcolor[HTML]{E8E8E8}0.648} &
  \multicolumn{1}{l|}{\cellcolor[HTML]{E8E8E8}0.933} &
  \multicolumn{1}{l|}{\cellcolor[HTML]{E8E8E8}0.986} &
  \multicolumn{1}{l|}{\cellcolor[HTML]{E8E8E8}1} &
  \multicolumn{1}{l|}{\cellcolor[HTML]{E8E8E8}0.986} &
  \multicolumn{1}{l|}{\cellcolor[HTML]{E8E8E8}0.947} &
  \multicolumn{1}{l|}{\cellcolor[HTML]{E8E8E8}0.866} &
  \multicolumn{1}{l|}{\cellcolor[HTML]{E8E8E8}0.65} &
  \cellcolor[HTML]{E8E8E8}0.374 \\ \cline{2-16} 
 &
  SMF &
  1310nm &
  \multicolumn{1}{l|}{0.003} &
  \multicolumn{1}{l|}{0.005} &
  \multicolumn{1}{l|}{0.012} &
  \multicolumn{1}{l|}{0.018} &
  \multicolumn{1}{l|}{0.023} &
  \multicolumn{1}{l|}{0.024} &
  \multicolumn{1}{l|}{0.021} &
  \multicolumn{1}{l|}{0.015} &
  \multicolumn{1}{l|}{0.004} &
  \multicolumn{1}{l|}{0} &
  \multicolumn{1}{l|}{0} &
  \multicolumn{1}{l|}{0} &
  0 \\ \cline{2-16} 
\multirow{-3}{*}{50} &
  \cellcolor[HTML]{D7F4F5}NZDSF &
  \cellcolor[HTML]{D7F4F5}1550nm &
  \multicolumn{1}{l|}{\cellcolor[HTML]{D7F4F5}0.011} &
  \multicolumn{1}{l|}{\cellcolor[HTML]{D7F4F5}0.039} &
  \multicolumn{1}{l|}{\cellcolor[HTML]{D7F4F5}0.18} &
  \multicolumn{1}{l|}{\cellcolor[HTML]{D7F4F5}0.401} &
  \multicolumn{1}{l|}{\cellcolor[HTML]{D7F4F5}0.609} &
  \multicolumn{1}{l|}{\cellcolor[HTML]{D7F4F5}0.745} &
  \multicolumn{1}{l|}{\cellcolor[HTML]{D7F4F5}0.782} &
  \multicolumn{1}{l|}{\cellcolor[HTML]{D7F4F5}0.791} &
  \multicolumn{1}{l|}{\cellcolor[HTML]{D7F4F5}0.774} &
  \multicolumn{1}{l|}{\cellcolor[HTML]{D7F4F5}0.735} &
  \multicolumn{1}{l|}{\cellcolor[HTML]{D7F4F5}0.656} &
  \multicolumn{1}{l|}{\cellcolor[HTML]{D7F4F5}0.453} &
  \cellcolor[HTML]{D7F4F5}0.215 \\ \hline
 &
  \cellcolor[HTML]{E8E8E8}SMF &
  \cellcolor[HTML]{E8E8E8}1550nm &
  \multicolumn{1}{l|}{\cellcolor[HTML]{E8E8E8}0.003} &
  \multicolumn{1}{l|}{\cellcolor[HTML]{E8E8E8}0.012} &
  \multicolumn{1}{l|}{\cellcolor[HTML]{E8E8E8}0.074} &
  \multicolumn{1}{l|}{\cellcolor[HTML]{E8E8E8}0.272} &
  \multicolumn{1}{l|}{\cellcolor[HTML]{E8E8E8}0.699} &
  \multicolumn{1}{l|}{\cellcolor[HTML]{E8E8E8}0.953} &
  \multicolumn{1}{l|}{\cellcolor[HTML]{E8E8E8}0.99} &
  \multicolumn{1}{l|}{\cellcolor[HTML]{E8E8E8}1} &
  \multicolumn{1}{l|}{\cellcolor[HTML]{E8E8E8}0.985} &
  \multicolumn{1}{l|}{\cellcolor[HTML]{E8E8E8}0.945} &
  \multicolumn{1}{l|}{\cellcolor[HTML]{E8E8E8}0.865} &
  \multicolumn{1}{l|}{\cellcolor[HTML]{E8E8E8}0.649} &
  \cellcolor[HTML]{E8E8E8}0.373 \\ \cline{2-16} 
 &
  SMF &
  1310nm &
  \multicolumn{1}{l|}{0.007} &
  \multicolumn{1}{l|}{0.012} &
  \multicolumn{1}{l|}{0.019} &
  \multicolumn{1}{l|}{0.023} &
  \multicolumn{1}{l|}{0.025} &
  \multicolumn{1}{l|}{0.024} &
  \multicolumn{1}{l|}{0.021} &
  \multicolumn{1}{l|}{0.015} &
  \multicolumn{1}{l|}{0.004} &
  \multicolumn{1}{l|}{0} &
  \multicolumn{1}{l|}{0} &
  \multicolumn{1}{l|}{0} &
  0 \\ \cline{2-16} 
\multirow{-3}{*}{5} &
  \cellcolor[HTML]{D7F4F5}NZDSF &
  \cellcolor[HTML]{D7F4F5}1550nm &
  \multicolumn{1}{l|}{\cellcolor[HTML]{D7F4F5}0.011} &
  \multicolumn{1}{l|}{\cellcolor[HTML]{D7F4F5}0.041} &
  \multicolumn{1}{l|}{\cellcolor[HTML]{D7F4F5}0.214} &
  \multicolumn{1}{l|}{\cellcolor[HTML]{D7F4F5}0.508} &
  \multicolumn{1}{l|}{\cellcolor[HTML]{D7F4F5}0.682} &
  \multicolumn{1}{l|}{\cellcolor[HTML]{D7F4F5}0.762} &
  \multicolumn{1}{l|}{\cellcolor[HTML]{D7F4F5}0.785} &
  \multicolumn{1}{l|}{\cellcolor[HTML]{D7F4F5}0.791} &
  \multicolumn{1}{l|}{\cellcolor[HTML]{D7F4F5}0.773} &
  \multicolumn{1}{l|}{\cellcolor[HTML]{D7F4F5}0.734} &
  \multicolumn{1}{l|}{\cellcolor[HTML]{D7F4F5}0.655} &
  \multicolumn{1}{l|}{\cellcolor[HTML]{D7F4F5}0.453} &
  \cellcolor[HTML]{D7F4F5}0.214 \\ \hline
\end{tabular}
    \caption{
    The normalised SKR per GHz of the bandwidth for a distance of \SI{100}{\kilo\meter} and many multiplexed channels.  
    Here the SKR is normalised to the maximum value seen in the combination of distance and jitter.
    }
    \label{tab:3.2.1-summary-100km}
\end{table}

\newpage
\begin{table}[ht]
\centering
\begin{tabular}{|l|l|l|lllllllllllll|}
\hline
 &
   &
   &
  \multicolumn{13}{l|}{Bandwidth, GHz} \\
\multirow{-2}{*}{Jitter} &
  \multirow{-2}{*}{Fibre} &
  \multirow{-2}{*}{Wavelength} &
  \multicolumn{1}{l|}{100} &
  \multicolumn{1}{l|}{50} &
  \multicolumn{1}{l|}{20} &
  \multicolumn{1}{l|}{10} &
  \multicolumn{1}{l|}{5} &
  \multicolumn{1}{l|}{2} &
  \multicolumn{1}{l|}{1} &
  \multicolumn{1}{l|}{0.5} &
  \multicolumn{1}{l|}{0.2} &
  \multicolumn{1}{l|}{0.1} &
  \multicolumn{1}{l|}{0.05} &
  \multicolumn{1}{l|}{0.02} &
  0.01 \\ \hline
 &
  \cellcolor[HTML]{E8E8E8}SMF &
  \cellcolor[HTML]{E8E8E8}1550nm &
  \multicolumn{1}{l|}{\cellcolor[HTML]{E8E8E8}0.016} &
  \multicolumn{1}{l|}{\cellcolor[HTML]{E8E8E8}0.037} &
  \multicolumn{1}{l|}{\cellcolor[HTML]{E8E8E8}0.097} &
  \multicolumn{1}{l|}{\cellcolor[HTML]{E8E8E8}0.192} &
  \multicolumn{1}{l|}{\cellcolor[HTML]{E8E8E8}0.363} &
  \multicolumn{1}{l|}{\cellcolor[HTML]{E8E8E8}0.687} &
  \multicolumn{1}{l|}{\cellcolor[HTML]{E8E8E8}0.872} &
  \multicolumn{1}{l|}{\cellcolor[HTML]{E8E8E8}0.956} &
  \multicolumn{1}{l|}{\cellcolor[HTML]{E8E8E8}0.992} &
  \multicolumn{1}{l|}{\cellcolor[HTML]{E8E8E8}1} &
  \multicolumn{1}{l|}{\cellcolor[HTML]{E8E8E8}1} &
  \multicolumn{1}{l|}{\cellcolor[HTML]{E8E8E8}0.992} &
  \cellcolor[HTML]{E8E8E8}0.975 \\ \cline{2-16} 
 &
  SMF &
  1310nm &
  \multicolumn{1}{l|}{0.014} &
  \multicolumn{1}{l|}{0.029} &
  \multicolumn{1}{l|}{0.071} &
  \multicolumn{1}{l|}{0.139} &
  \multicolumn{1}{l|}{0.25} &
  \multicolumn{1}{l|}{0.412} &
  \multicolumn{1}{l|}{0.48} &
  \multicolumn{1}{l|}{0.506} &
  \multicolumn{1}{l|}{0.516} &
  \multicolumn{1}{l|}{0.517} &
  \multicolumn{1}{l|}{0.516} &
  \multicolumn{1}{l|}{0.507} &
  0.492 \\ \cline{2-16} 
\multirow{-3}{*}{200} &
  \cellcolor[HTML]{D7F4F5}NZDSF &
  \cellcolor[HTML]{D7F4F5}1550nm &
  \multicolumn{1}{l|}{\cellcolor[HTML]{D7F4F5}0.019} &
  \multicolumn{1}{l|}{\cellcolor[HTML]{D7F4F5}0.038} &
  \multicolumn{1}{l|}{\cellcolor[HTML]{D7F4F5}0.096} &
  \multicolumn{1}{l|}{\cellcolor[HTML]{D7F4F5}0.188} &
  \multicolumn{1}{l|}{\cellcolor[HTML]{D7F4F5}0.356} &
  \multicolumn{1}{l|}{\cellcolor[HTML]{D7F4F5}0.672} &
  \multicolumn{1}{l|}{\cellcolor[HTML]{D7F4F5}0.852} &
  \multicolumn{1}{l|}{\cellcolor[HTML]{D7F4F5}0.935} &
  \multicolumn{1}{l|}{\cellcolor[HTML]{D7F4F5}0.97} &
  \multicolumn{1}{l|}{\cellcolor[HTML]{D7F4F5}0.977} &
  \multicolumn{1}{l|}{\cellcolor[HTML]{D7F4F5}0.977} &
  \multicolumn{1}{l|}{\cellcolor[HTML]{D7F4F5}0.969} &
  \cellcolor[HTML]{D7F4F5}0.952 \\ \hline
 &
  \cellcolor[HTML]{E8E8E8}SMF &
  \cellcolor[HTML]{E8E8E8}1550nm &
  \multicolumn{1}{l|}{\cellcolor[HTML]{E8E8E8}0.025} &
  \multicolumn{1}{l|}{\cellcolor[HTML]{E8E8E8}0.077} &
  \multicolumn{1}{l|}{\cellcolor[HTML]{E8E8E8}0.249} &
  \multicolumn{1}{l|}{\cellcolor[HTML]{E8E8E8}0.465} &
  \multicolumn{1}{l|}{\cellcolor[HTML]{E8E8E8}0.69} &
  \multicolumn{1}{l|}{\cellcolor[HTML]{E8E8E8}0.879} &
  \multicolumn{1}{l|}{\cellcolor[HTML]{E8E8E8}0.946} &
  \multicolumn{1}{l|}{\cellcolor[HTML]{E8E8E8}0.978} &
  \multicolumn{1}{l|}{\cellcolor[HTML]{E8E8E8}0.996} &
  \multicolumn{1}{l|}{\cellcolor[HTML]{E8E8E8}1} &
  \multicolumn{1}{l|}{\cellcolor[HTML]{E8E8E8}1} &
  \multicolumn{1}{l|}{\cellcolor[HTML]{E8E8E8}0.991} &
  \cellcolor[HTML]{E8E8E8}0.974 \\ \cline{2-16} 
 &
  SMF &
  1310nm &
  \multicolumn{1}{l|}{0.043} &
  \multicolumn{1}{l|}{0.085} &
  \multicolumn{1}{l|}{0.194} &
  \multicolumn{1}{l|}{0.311} &
  \multicolumn{1}{l|}{0.41} &
  \multicolumn{1}{l|}{0.479} &
  \multicolumn{1}{l|}{0.501} &
  \multicolumn{1}{l|}{0.512} &
  \multicolumn{1}{l|}{0.517} &
  \multicolumn{1}{l|}{0.517} &
  \multicolumn{1}{l|}{0.515} &
  \multicolumn{1}{l|}{0.507} &
  0.492 \\ \cline{2-16} 
\multirow{-3}{*}{50} &
  \cellcolor[HTML]{D7F4F5}NZDSF &
  \cellcolor[HTML]{D7F4F5}1550nm &
  \multicolumn{1}{l|}{\cellcolor[HTML]{D7F4F5}0.05} &
  \multicolumn{1}{l|}{\cellcolor[HTML]{D7F4F5}0.108} &
  \multicolumn{1}{l|}{\cellcolor[HTML]{D7F4F5}0.264} &
  \multicolumn{1}{l|}{\cellcolor[HTML]{D7F4F5}0.463} &
  \multicolumn{1}{l|}{\cellcolor[HTML]{D7F4F5}0.677} &
  \multicolumn{1}{l|}{\cellcolor[HTML]{D7F4F5}0.86} &
  \multicolumn{1}{l|}{\cellcolor[HTML]{D7F4F5}0.924} &
  \multicolumn{1}{l|}{\cellcolor[HTML]{D7F4F5}0.956} &
  \multicolumn{1}{l|}{\cellcolor[HTML]{D7F4F5}0.973} &
  \multicolumn{1}{l|}{\cellcolor[HTML]{D7F4F5}0.977} &
  \multicolumn{1}{l|}{\cellcolor[HTML]{D7F4F5}0.977} &
  \multicolumn{1}{l|}{\cellcolor[HTML]{D7F4F5}0.968} &
  \cellcolor[HTML]{D7F4F5}0.952 \\ \hline
 &
  \cellcolor[HTML]{E8E8E8}SMF &
  \cellcolor[HTML]{E8E8E8}1550nm &
  \multicolumn{1}{l|}{\cellcolor[HTML]{E8E8E8}0.026} &
  \multicolumn{1}{l|}{\cellcolor[HTML]{E8E8E8}0.089} &
  \multicolumn{1}{l|}{\cellcolor[HTML]{E8E8E8}0.346} &
  \multicolumn{1}{l|}{\cellcolor[HTML]{E8E8E8}0.599} &
  \multicolumn{1}{l|}{\cellcolor[HTML]{E8E8E8}0.768} &
  \multicolumn{1}{l|}{\cellcolor[HTML]{E8E8E8}0.899} &
  \multicolumn{1}{l|}{\cellcolor[HTML]{E8E8E8}0.951} &
  \multicolumn{1}{l|}{\cellcolor[HTML]{E8E8E8}0.979} &
  \multicolumn{1}{l|}{\cellcolor[HTML]{E8E8E8}0.996} &
  \multicolumn{1}{l|}{\cellcolor[HTML]{E8E8E8}1} &
  \multicolumn{1}{l|}{\cellcolor[HTML]{E8E8E8}1} &
  \multicolumn{1}{l|}{\cellcolor[HTML]{E8E8E8}0.991} &
  \cellcolor[HTML]{E8E8E8}0.974 \\ \cline{2-16} 
 &
  SMF &
  1310nm &
  \multicolumn{1}{l|}{0.099} &
  \multicolumn{1}{l|}{0.169} &
  \multicolumn{1}{l|}{0.288} &
  \multicolumn{1}{l|}{0.373} &
  \multicolumn{1}{l|}{0.436} &
  \multicolumn{1}{l|}{0.485} &
  \multicolumn{1}{l|}{0.503} &
  \multicolumn{1}{l|}{0.512} &
  \multicolumn{1}{l|}{0.517} &
  \multicolumn{1}{l|}{0.517} &
  \multicolumn{1}{l|}{0.515} &
  \multicolumn{1}{l|}{0.507} &
  0.492 \\ \cline{2-16} 
\multirow{-3}{*}{5} &
  \cellcolor[HTML]{D7F4F5}NZDSF &
  \cellcolor[HTML]{D7F4F5}1550nm &
  \multicolumn{1}{l|}{\cellcolor[HTML]{D7F4F5}0.072} &
  \multicolumn{1}{l|}{\cellcolor[HTML]{D7F4F5}0.187} &
  \multicolumn{1}{l|}{\cellcolor[HTML]{D7F4F5}0.426} &
  \multicolumn{1}{l|}{\cellcolor[HTML]{D7F4F5}0.605} &
  \multicolumn{1}{l|}{\cellcolor[HTML]{D7F4F5}0.754} &
  \multicolumn{1}{l|}{\cellcolor[HTML]{D7F4F5}0.879} &
  \multicolumn{1}{l|}{\cellcolor[HTML]{D7F4F5}0.93} &
  \multicolumn{1}{l|}{\cellcolor[HTML]{D7F4F5}0.957} &
  \multicolumn{1}{l|}{\cellcolor[HTML]{D7F4F5}0.973} &
  \multicolumn{1}{l|}{\cellcolor[HTML]{D7F4F5}0.977} &
  \multicolumn{1}{l|}{\cellcolor[HTML]{D7F4F5}0.977} &
  \multicolumn{1}{l|}{\cellcolor[HTML]{D7F4F5}0.968} &
  \cellcolor[HTML]{D7F4F5}0.952 \\ \hline
\end{tabular}
    \caption{
    The normalised SKR per GHz of the bandwidth for a distance of \SI{10}{\kilo\meter} and many multiplexed channels. 
    Here the SKR is normalised to the maximum value seen in the combination of distance and jitter.
    }
    \label{tab:3.2.1-summary-10km}
\end{table}

\end{landscape}

\newpage


 \newcommand{\noop}[1]{}


\end{document}